ORIGINAL ARTICLE

# Transformer-based time series prediction of the maximum power point for solar photovoltaic cells

**Palaash Agrawal[1]**  | **Hari Om Bansal[1]**  | **Aditya R. Gautam[1]**  |
**Om Prakash Mahela[2]** | **Baseem Khan[3]**

[1]Birla Institute of Technology and Science, Pilani, Rajasthan, India

[2]Power System Planning Division, Rajasthan Rajya Vidyut Prasaran Nigam Ltd., Jaipur, Rajasthan, India

[3]Department of Electrical and Computer Engineering, Hawassa University, Hawassa, Ethiopia

**Correspondence**

Baseem Khan, Department of Electrical and computer engineering, Hawassa University, Hawassa, Ethiopia.
Email: baseem.khan04@ieee.org

**Abstract**

This paper proposes an improved deep learning-based maximum power point tracking (MPPT) in solar photovoltaic cells considering various time series-based environmental inputs. Generally, artificial neural network-based MPPT algorithms use basic neural network architectures and inputs which do not represent the ambient conditions in a comprehensive manner. In this article, the ambient conditions of a location are represented through a comprehensive set of environmental features. Furthermore, the inclusion of time-based features in the input data is considered to model cyclic patterns temporally within the atmospheric conditions leading to robust modeling of the MPPT algorithm. A transformer-based deep learning architecture is trained as a time series prediction model using multidimensional time series input features. The model is trained on a dataset containing typical meteorological-year data points of ambient weather conditions from 50 locations. The attention mechanism in the transformer modules allows the model to learn temporal patterns in the data efficiently. The proposed model achieves a 0.47% mean average percentage error of prediction on non-zero operating voltage points in a test dataset consisting of data collected over a period of 200 consecutive hours; resulting in the average power efficiency of 99.54% and peak power efficiency of 99.98%. The proposed model is validated through real-time simulations. The proposed model performs power point tracking in a robust, dynamic, and nonlatent manner, over a wide range of atmospheric conditions.

## 1 | INTRODUCTION

Following the trend of increasing energy demands, and the possibility of exhaustion of all fossil fuel-based sources, the focus is rapidly shifting towards renewable energy sources. One of the most promising avenues in this domain is solar energy. Extensive research is being conducted to improve solar panel efficiency and in designing modules that can be integrated into various physical structures. Solar photovoltaic (PV) cells can now be installed not only in fields and rooftops, but as solar trees, floating systems, building facades, and even automobile vehicles.[1,2] However, harnessing solar energy is a challenging task since solar irradiation levels fluctuate to a great extent. Solar irradiation varies by many environmental factors,[3] such as season, natural ambient conditions, man-made conditions such as air pollution,[4] and phenomena such as dust deposition on solar panels,[5] or partial shading.[6]

One of the biggest issues with harnessing solar energy is its dilute nature,[7] which implies that to compensate for the varying atmospheric conditions across the globe, the surface area required to produce the same amount of energy varies greatly, given that all other factors remain constant. Moreover, given the limited nature of energy conversion efficiency in PV cells, methods have to be employed to harness the maximum amount of energy. Significant research has been carried out in the field of maximum power point tracking (MPPT) over the past decades. This includes various techniques including perturb and observe (P&O), hill climbing, incremental conductance, short-circuit current, open-circuit voltage, ripple correlation, fuzzy logic, particle swarm optimization, and so forth. Apart from these methods, various heuristic methods have also been used to perform MPPT in environments such as partial shading conditions.[8]

Among various optimization methods, deep learning-based methods have demonstrated the highest accuracy, flexibility, and robustness. Various neural network-based approaches have been proposed in the last decade. It usually consists of a three-layer fully connected neural network which is fed by physical inputs measurable by small sensors.[9,10] Other approaches towards artificial neural network (ANN)-based modeling of MPPT involve the application of radial basis functions (RBF),[11] fuzzy logic operators,[12] genetic algorithms,[13] heuristic search-based algorithms,[14] and a combination with traditional MPPT techniques such as P&O[15] and incremental conductance.[16] Apart from the traditional fully connected ANN approach, some works have attempted the MPPT task using approaches such as reinforcement learning (RL)[17] and classical machine learning approaches[18] such as decision trees, k-nearest neighbors, and recurrent neural networks (RNNs). For tracking MPP in PV systems, inputs to ANN-based models can majorly be categorized into two broad classes namely, environmental inputs—temperature ($T$) and irradiance ($G$) and electrical inputs current and voltage.[19,20]

Giraud and Salameh[21] proposed the effects of passing clouds on a grid-tied PV system with battery storage using combined RBF and a backpropagation network.

However, only irradiance has been used as input to model for environmental variations. The authors consider cloud movement as the factor that influences climate which alone is insufficient to characterize ambient conditions since many other factors such as the relative solar altitude angle which directly affect surface irradiance along with cloud cover.[22] In the study by Ramaprabha,[23] an improved backpropagation neural network-based MPPT for boost converter is proposed which uses irradiance and temperature conditions as inputs to predict the MPP of solar PV. The predicted operating voltage and current values are used to continuously adjust the power point of the PV cells. However, this basic implementation is designed for a small range of environmental conditions. Zhang et al.[24] proposed a more sophisticated architecture that utilizes a genetic algorithm optimized RBF-based ANN. Moreover, the cell temperature is used as the input, which is assumed to represent the collective effect of various fundamental climatic factors including wind speed and ambient temperature. However, the effect of delayed latent heat dissipation is not considered, meaning that cell temperature is not a dynamic representation of climatic conditions. In case of sudden ambient temperature or wind speed fluctuation, the cell temperature fails to change at the same rate because of the physical properties of the PV cell. A better method would be to consider the more fundamental climatic factors as inputs themselves. In the study by Chaouachi et al.,[25] fuzzy logic is employed, as a result of which, temperature and irradiance are classified into three categories each, effectively making the input range to nine values, which is insufficient to represent the vast spectrum of ambient conditions. Some other input types include error signal (defined as $E = (P_k - P_{k-1})/(V_k - V_{k-1})$) and error rate (defined as $\Delta E = E_k - E_{k-1}$).[26,27] Inputs as a function of time have also been modeled in a few works. In various studies,[28-30] the hour of the day is fed into the model as an input. However, this parameter itself is insufficient to represent short-term weather patterns over multiple geographical locations since it would not be possible to represent the different ambient conditions in different locations through a single time-based parameter. Open-circuit voltage has also been used as a complementary input but has a direct linear correlation with the operating voltage, which is a source of data leakage. Finally, the general approach in the domain of ANN-based MPPT is to feed data points as independent inputs, thus leading to the incapacity of neural networks to identify temporal patterns.

Previous ANN-based MPPT models mostly use basic feed-forward fully connected networks. While fully connected ANN models have demonstrated good performance,[31] however, they cannot generalize well over long-term environmental variations, due to their inability to model extremely complex relations. Moreover, nowadays several advanced architectures are available which can model complex relations and perform better than basic fully connected ANN models. Furthermore, basic ANN architectures treat each time instance independently which leads to a lack of exploitation in the cyclic nature of environmental variations in the short as well as long term. A better approach would be to use a time-series architecture, that can efficiently model long-term dependencies and patterns. Some other neural network-based architectures include RL-based methods.[32] This technique relies on the exploration of parameter values, which allows higher dynamicity in predictions.[33] In the study by Yadav and Chowdhury,[34] a deep-Q learning-based RL model was proposed to predict the fractional increment in operating voltage based on electrical parameters. This model was tested on data representing very limited environmental variations and its effectiveness in mapping the inputs with the MPP in diverse climatic conditions was not explored. Also, RL models are hard to optimize with the increasing size of the underlying neural network model. Hence, pure ANN-based approaches outperform RL-based approaches in tracking MPP. ANNs are also capable of generalizing over diverse climatic conditions, which has not yet been demonstrated by RL-based techniques. Another neural network approach is the RNN.[18] RNNs are usually suitable for time series tasks. However, the authors have used RNNs as standard deep learning models, to which data is fed batch-wise, thus preventing the model from learning the temporal patterns. This has resulted in poor performance of recurrent neural-based MPPT approaches.

In this paper, MPP tracking is considered a time series task. This is because of a temporal pattern in the solar cell characteristics, which arises because of a cyclic pattern in the ambient weather conditions. Moreover, the ambient weather conditions are represented through a comprehensive set of features, which include various factors that affect the performance of a solar PV cell. In addition, two more time-based features have been proposed; one represents the daily variations in the solar cycle (the day/night cycle) and the other represents the seasonal variations. These features help the model in effectively identifying the cyclic pattern in weather conditions, which forms the backbone of the time series prediction task. Here, a powerful transformer-based architecture is used which can effectively learn long-range time dependencies thus making it suitable for the MPP prediction.[35]

The major contributions of this paper are as follows:

1. A novel set of data input features based on environmental factors is proposed, that can extensively model the true ambient conditions at any given time.
2. Within this set of features, a set of discrete time-based inputs is proposed, representing short-term (daily) and long-term (seasonal) variation in the ambient conditions, which helps the model identify the cyclic nature of ambient conditions, and hence the resulting cyclic nature of the operating point of the solar PV cell.
3. A transformer-based architecture is employed to develop a multidimensional time series prediction model. This transformer architecture is optimized through a mixture of various state-of-the-art like a cyclic learning rate and momentum hyperparameter schedulers, and optimal learning rate value analysis, that have not been considered in previous work in this domain.

The remainder of the paper is organized as follows. Section 2 presents the materials used in the experiments. Section 3 describes the proposed modeling technique and architecture elaborating the exact structuring and processing methods applied to the data. It also details on the training of the model and mathematical justifications of various architectural choices. The results of the proposed method are described in Section 4, and various ablation studies along with details on hardware implementation and comparison of the performance with related works have been presented in Section 5. Finally, concluding remarks are given in Section 6, where the work presented in this paper has been summarized, and future research directions have been discussed.

## 2 | MATERIALS AND METHODS

In this section, the proposed set of data features along with supporting experiments, studies, and analyses have been presented.

### 2.1 | Data selection

The efficiency of a solar PV cell varies on various ambient factors, directly, or indirectly. Traditional parameters such as temperature and solar irradiation do not completely characterize the atmospheric condition of a location. Two locations with the same solar irradiation and ambient temperature can yield different PV operating points, owing to other factors such as humidity, wind, dust deposition, cloud cover, and so forth. Thus, a more inclusive selection of atmospheric parameters is needed. The aim is to choose independent parameters that have a direct and significant impact on solar cell efficiency. Some of these features have been mentioned below to justify their importance.

### 2.1.1 | Ambient temperature

An increase in ambient temperature causes an increase in short-circuit current and a significant drop in open-circuit voltage. As a result, the PV cell's overall power output decreases. Figure 1A describes the dependence of PV performance on the ambient temperature. The black dots in the figure represent the MPP of the cell at a given temperature.

### 2.1.2 | Solar irradiation

An increase in the irradiation intensity falling on a PV cell increases its power output. The photocurrent produced increases, and thus the operating voltage by a relatively smaller amount. This dependence of PV performance with solar irradiation is represented in Figure 1B. Irradiation can be divided in two components direct normal irradiance (DNI) and diffuse horizontal irradiance (DHI). Both these features are considered in the proposed work.

### 2.1.3 | Time-based inputs

Ambient conditions demonstrate repeated patterns over short as well as longer periods of time (see Figure 2). This repetition in the pattern is termed weather in the short-term (daily) domain and seasonality in the long-term domain. The cyclic nature of ambient environmental conditions implies that solar cell performance would demonstrate this repeated pattern too. Hence, we represent the long- and short-term temporal patterns through discrete time-based representations.

*Month-of-the-year*
Long-term seasonal behavior is represented and discretized in 12 integral steps representing the variation that occurs within a month. The month of the year is chosen as the discretization basis due to its intuitive nature and more-or-less uniform distribution of seasonal characteristics across the various months. Using this input, the ANN can also identify variation in climatic and solar patterns along the longitude of the earth, that is, from one pole of the earth to the other, by establishing a suitable mathematical relation among month-of-the-year

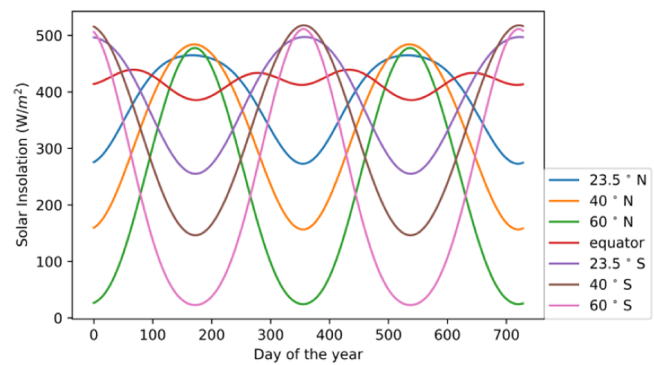

**FIGURE 2** Qualitative variation of peak solar radiation energy with latitudinal change over two years.

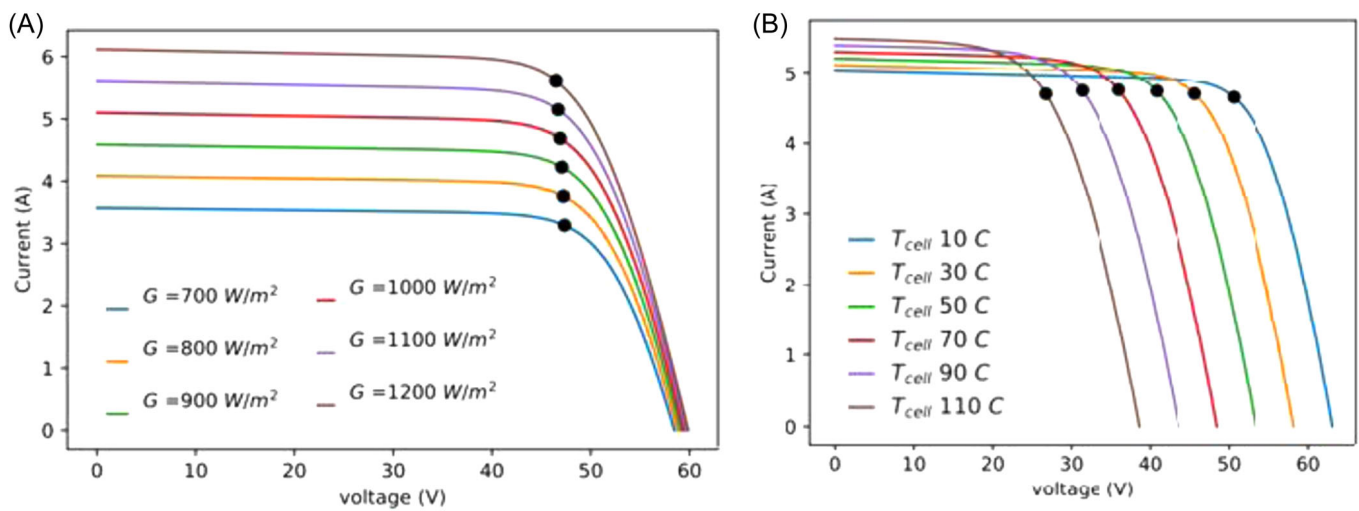

**FIGURE 1** *I–V* characteristics. (A) Temperature (*T*) dependence at constant irradiance $G = 1000 \, \text{W/m}^2$. (B) Irradiance (*G*) dependence at constant temperature $T = 25°C$

input, solar azimuth angle, and other inputs that may reflect related patterns of seasonal variation.

*Hour-of-the-day*

Short-term behavior is represented and discretized in the form of an hour of the day (24-step discretization). This input represents overall cyclic trends in the power output resulting from climatic and solar variations during the entire daytime. This parameter also accounts for the change in properties of solar irradiation throughout the day. One particularly important property is the band of wavelengths of sunlight reaching the ground. Solar PV cell efficiency peaks for a particular wavelength,[36] and thus, is expected to provide the maximum power output at a given time of the day. Hence, the neural network can potentially identify seasonal trends in the daily climatic/solar cycle as well, by establishing suitable mathematical relations between hour-of-the-day, solar altitude angle, the solar azimuth angle, and any other parameters which have related patterns.

The various other features considered here are solar altitude angle,[37] solar azimuth angle,[38] wind speed,[39,40] humidity,[41–43] and pollution level (PM10 concentration).[5,44] Through the combination of the various features proposed above, the deep learning model is able to correlate the various repeating patterns, ensuring high accuracy in the prediction results.

## 2.2 | Dataset specifications

For this model, data has been collected for 50 cities in India, available from the USA National Renewable Energy Laboratory's SUNY Semi-Empirical Satellite Model developed by Perez et al. as part of the India–U.S. Energy Dialogue.[45–47] Pollution levels are collected from the ambient air quality database maintained by the Centeral Pollution Control Board (CPCB). The weather data is simulated on system advisor model's photovoltaic performance model,[48] to obtain maximum power output, operating voltage, and operating current, for a standard crystalline silicon PV module. The details of the crystalline silicon PV module are given in Table 1.

The model has been trained to predict operating voltage, whose product with corresponding operating current from the ground data is compared with the maximum power output. While this model has been designed to predict the operating voltage directly, a similar model can be used to predict the voltage duty cycle instead. Alternatively, duty cycle parameters can be derived directly from the predicted output voltage to drive the MPP power converter.

**TABLE 1** PV module specifications.

| Parameter | Value |
|---|---|
| Approximate nominal efficiency | 19% |
| Module cover | Antireflective glass |
| Temperature coefficient of power | −0.37%/°C |
| Fill factor (for self-shading) | 77.8% |
| Nominal power capacity | 0.23 kWdc |
| Rated inverter size | 0.20 kWac |
| Inverter efficiency | 96% |
| Array Type | Fixed roof mount |
| Tilt | 20° |
| Azimuth | 180° |
| Ground coverage ratio | 0.3 |
| Total module area | 4.035 $m^2$ |

Abbreviation: PV, photovoltaic.

## 3 | PROPOSED MODELING TECHNIQUE AND ARCHITECTURE

This section discusses data categorization, model architecture, and modeling techniques.

### 3.1 | Data categorization and representation

Neural networks can take in various types of inputs, provided it is first converted to an appropriate format. The simplest form of input is a real integer/float belonging to a continuous distribution, which can be directly fed through the neural network as separate features. The mathematical derivative can also be easily obtained through backpropagation. Such inputs are categorized as continuous variables which include ambient temperature, DNI, DHI, solar altitude angle, solar azimuth angle, wind speed, humidity, and pollution levels. They can be fed into the neural network as floating-point numbers. On the contrary, inputs such as month-of-the-year and hour-of-the-day are variables whose values lie in either of a fixed number of categories and thus cannot be treated like continuous variables. Such variables are categorized under categorical or ordinal variables. The categorization of the features used is depicted in Figure 3A,B represents the categorical or ordinal variables fed into the neural network as one-hot-encoded matrices. In Figure 3B, the hours of the day have been represented in 24-h format.

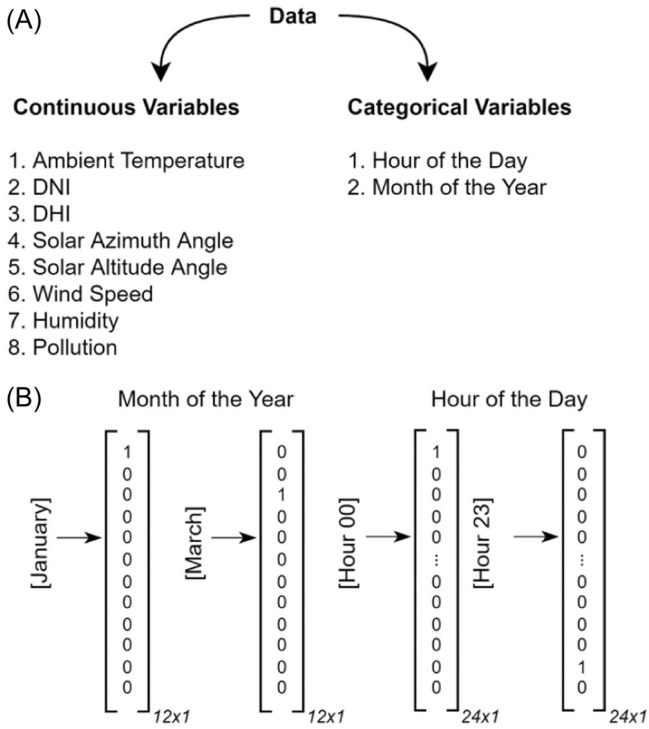

**FIGURE 3** (A) Data visualization, (B) categorical variables as one-hot-encoded matrices

## 3.2 | Entity embedding for categorical variables

A one-hot-encoded matrix, when propagated through the neural network by multiplication with parameter matrices, is essentially a series of matrix multiplications with most of the computation consisting of zero multiplications. Moreover, one-hot encodings are very sparse in information. Hence, to overcome this problem, categorical variables are first fed into entity embeddings transformations and then fed into neural network models.[49,50] In the proposed work, entity embeddings are linearly transformed representations of data, where each element of the input one-hot-encoded matrix is represented as a vector of $n$ features. The matrix elements act as parameters for the neural network and can be optimized to fit the target output, through standard backpropagation.

Consider the two-dimensional vector matrix as $u_{n \times m}$, where $n$ is the cardinality of the input one-hot-encoded matrix and $m$ is the embedding size, or in other terms the size of the feature space that acts as input to the following layers of the model. This matrix is called the embedding matrix. On the dot-product of the embedding matrix with the input one-hot-encoded matrix, a particular row of the embedding matrix is obtained. This row vector represents m features corresponding to a given input. Let input one-hot encoded vector of shape $1 \times n$ be represented by $i_k$, where the $k$th element of the one-hot encoded vector $i$ is unity, and the rest $n - 1$ elements are zeroes. Then, the dot product gives the following equation:

$$u_k = i_k \cdot u, \tag{1}$$

of order $1 \times m$ where $u_k$ is the $k$th column of the embedding matrix $u$, and is essentially a vector of size $m$. Hence, all 12 different inputs corresponding to the *month-of-the-year* have 12 different sets of $m$ features, and so is the case for the variable *hour-of-the-day*. Entity embeddings create arbitrary features of the input space without any specific feature engineering. Each input is mapped to a feature space of $m$ dimensions, and these can be used to deduce highly complex patterns. Hence, embeddings are particularly useful for time-based data, which do not represent any particular property or phenomenon but are generic in nature.

## 3.3 | Data preprocessing

First, all continuous variables are normalized independently, using a min–max scheme, and the categorical variables are tokenized. The categorical and continuous variables are separated and converted to the corresponding input vectors. Both of them are then converted into time series data points, which include stacking the current data point $X_T$ with $T - 1$ preceding data points collected from the same location. For the purpose of our model, $T$ is chosen 50 as it is large enough to capture short-term local weather shifts (spanning over a few days) but is small enough to avoid capturing long term (such as entire seasons) weather shifts, causing feature leakage.

Hence, each data point is represented as a combination (tuple) of continuous ($x_{\text{cont}}$) and categorical ($x_{\text{cat}}$) variables, where $x_{\text{cont}}$ has a shape of $T \times p$, where $p$ is the number of categorical features, and $x_{\text{cat}}$ has a shape of $T \times N$, where $N$ is the sum of the lengths of all one-hot encoded vectors as in Equation (2).

$$N = \sum_{i}^{X_{\text{cat}}} n^i. \tag{2}$$

To accelerate the data optimization process, the data is divided into segments or batches of size $B$ each, and the model learns by training on only one batch at once. This batch may contain random time series data points, thus resulting in ($x_{\text{cont}}$, $x_{\text{cat}}$) having the array dimensions ($B \times T \times p$, $B \times T \times N$). $B$ can be chosen according to the GPU capacity. In our case, a batch size of $B = 128$ was used.

## 3.4 | Architecture of the model

The model used is similar to that presented by Zerveas et al.[51] However, certain changes are proposed to handle two different forms of inputs. From each batch of the data, the continuous and categorical variables are separated. The two categorical variables are initially one-hot-encoded, then concatenated, and thereafter corresponding embedding feature space is looked up, which results in embedding matrix of dimension $N \times m$. The continuous variables (after normalization) are first projected onto a $d$-dimensional vector space (known as the *model dimension*). This allows the mapping process from predefined data features to homogenized neural network inputs to be totally learnable.

The two types of inputs undergo similar matrix transformations and are independent of each other. Hence the transformation process can be simplified by first concatenating the continuous variables and categorical one-hot encoded features and applying the projection function on all of them simultaneously to obtain a $T \times d$-dimensional feature. Hence, $m$ is replaced by $d$ in the embedding matrix $u$, resulting in the matrix dimensions of $N \times d$. The projection function dimension $d$ must be adjusted according to the size of the concatenated input to yield the adjusted model dimension value. The effective model dimension value is henceforth referred to as $d_{\text{eff}}$, which corresponds to the input size of the projection layer for continuous variables, and the adjusted model dimensions input size (including dimensions corresponding to zero values from categorical variables) is referred to as $d_{\text{adj}}$. The relation between $d_{\text{adj}}$ and $d_{\text{eff}}$ is given by Equation (3).

$$d_{\text{adj}} = d_{\text{eff}} + N, \tag{3}$$

which leads to the projection layer having the dimensions $d_{\text{adj}} \times d$.

In this paper, the encoder of the original transformer architecture[52] is used as the architecture for training. First, a fully learnable positional encoding is applied to the input encodings, followed by a padding mask. The data is then fed into the self-attention layer of the transformer encoder, after which it undergoes batch-normalization, a feed-forward layer, followed by another batch-normalization layer. The model utilizes three identical encoder blocks connected in a feed-forward fashion. Finally, another fully connected linear neuron layer with a single output (i.e., predicted voltage values) is attached. A sigmoid function on top of the predicted value is used to ensure the value remains within reasonable limits. The scaled sigmoid function ranges between the statistical maximum and minimum values of the operating voltage as in the training data. The final output $\widehat{y_i'}$ corresponding to the $i$th input element $x_i$ can be modeled as Equation (4). The architecture is depicted in Figure 4.

$$\widehat{y_i'} = f(\widehat{y_i}, y_i) = \left\{ [\max(y_i) - \min(y_i)] \times \frac{1}{1 + e^{-\widehat{y_i}}} + \min(y_i) \right\}. \tag{4}$$

## 3.5 | Training method

The data is first split into training and testing datasets. For this, a city is selected at random, and from it is separated out the last 200 h of data for the purpose of testing. The remaining data is used as training data. Since this is a time series problem, the data is not shuffled randomly. Each input data point is preprocessed through a min–max normalization, followed by concatenation into a time series array of 50 consecutive data points, where each data point itself is multivariate in nature. However, randomly picked and preprocessed time series data arrays can be batched randomly while training the deep learning model.

A pretrained architecture outperforms untrained models by a huge margin. Hence, initially, a transformer architecture pretrained on various time series tasks is used here. Then the architecture is fine-tuned using the Adam Optimizer[53] with the default hyperparameters and the one-cycle optimization strategy[54,55] which uses a continuous learning rate and momentum scheduler. The learning rate scheduler ranges from $10^{-6}$ to $10^{-2}$. From the plot of the training cycle with exponentially growing learning rates, the obtained range of peak learning rate is from $10^{-3}$ to $3 \times 10^{-2}$. The range with the steepest and smoothest decline in loss values was used. Similarly, momentum has a peak value of 0.1 and ranges from 0.01 to 0.1 throughout the hyperparameter cycle of training. All other hyperparameters are kept constant throughout the training cycle. The loss function used is a standard mean-squared error loss. Dropout is also used with a probability of neuron inactivation equal to 0.2. The fine-tuning process was carried out for 50 epochs and took anywhere from 8 to 12 h to complete training during multiple runs.

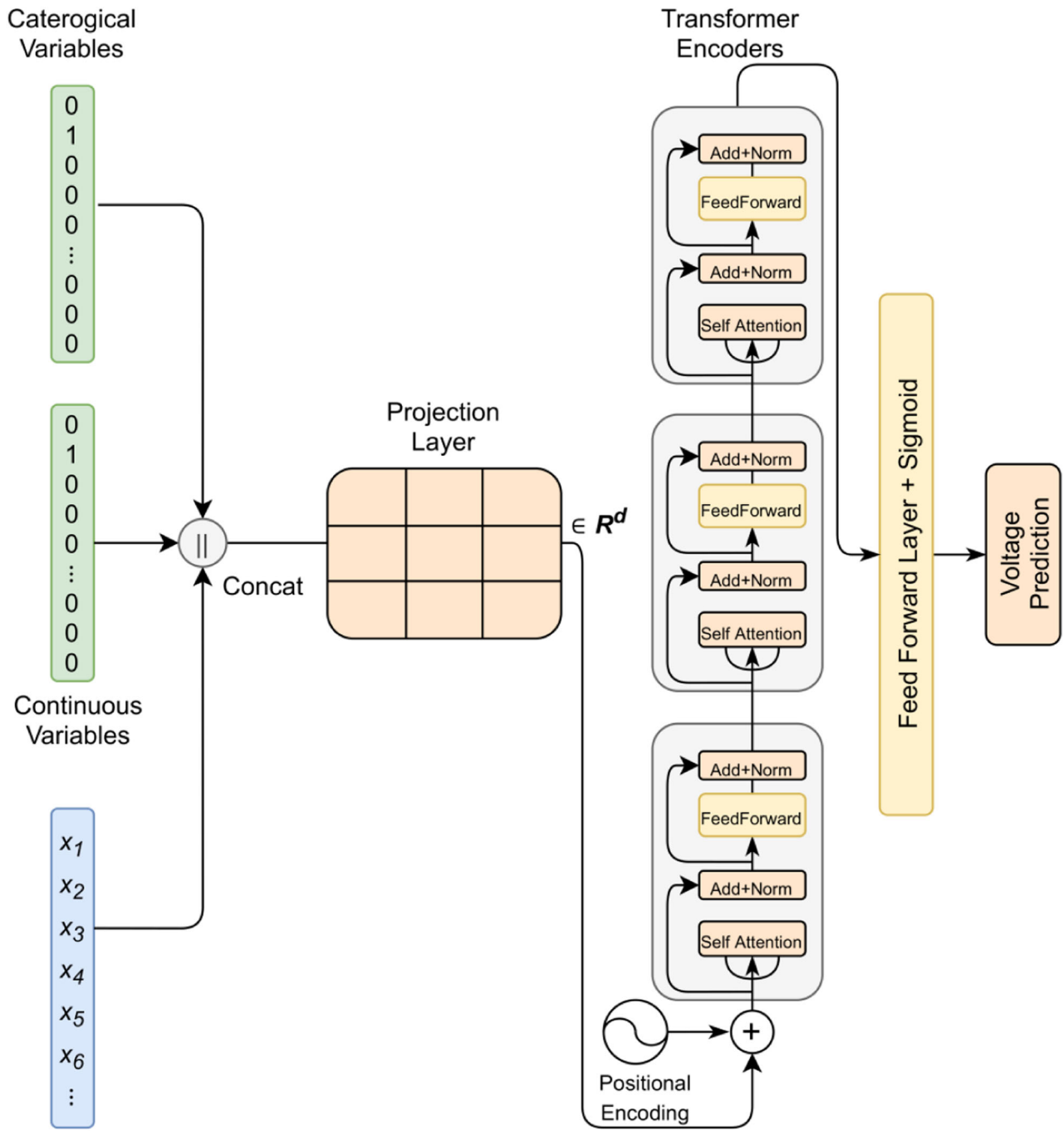

**FIGURE 4** Architecture of the proposed model

## 4 | RESULTS

This section discusses the simulation and real-time hardware results using proposed model. For inference, the weights are multiplied with the corresponding retention probabilities "$p$" (i.e., 1—dropout value). This is done to compensate for the extra weights present during testing, which were not present during training. The test output can be expected to be the same as obtained during training. The test dataset is chosen as the last 200 consecutive hours of data points

(each processed as a time series input consisting of 50 consecutive datapoints) were selected from a random city. The input data was forward propagated through the neural network. The resulting operating voltage predicted by the model is multiplied by the corresponding current derived from the simulated $I$–$V$ curve of the PV cell at the given temperature and irradiance values, simulated through the PVlib library. The resulting power output is compared to the ground truth MPP corresponding to the climatic/solar conditions.

The model yields a mean absolute percentage (MAP) error of 0.47% on the prediction of non-zero operating voltage points. The resulting operating power point, which is calculated by multiplying the predicted operating voltage with the corresponding current obtained from $I$–$V$ characteristics, results in an average MPPT efficiency (ratio of predicted power to actual power, i.e., MPP ratio) of 99.54%. At its peak, the model achieves an MPP ratio of 99.98%, thus proving the effectiveness of the model. Besides that, the neural network can follow the curvature of the actual data with a satisfactory amount of sharpness and nonlatency as shown in Figure 5.

For the purpose of this experiment, the OPAL-RT Simulator was used, which uses a Simulink-based model implementation to process input. These devices have FPGA-based computational engines, replicate real-world operations, and control the system effectively. The system performance is captured on a mixed signal oscilloscope (MSO). Figure 5A depicts the actual cell current corresponding to MPP at the given time instance for ambient conditions and test dataset. Figure 5B shows the ground truth operating voltage for the given PV cell, whereas Figure 5C shows the operating voltage predicted by the model. Corresponding to the ground truth voltage at the given time instance, the ground truth operating power for the cell is plotted in Figure 5D. This is calculated by multiplying the ground truth operating current with the ground truth voltage value. Finally, Figure 5E plots the power predicted by the model. Figure 6 shows the experimental setup.

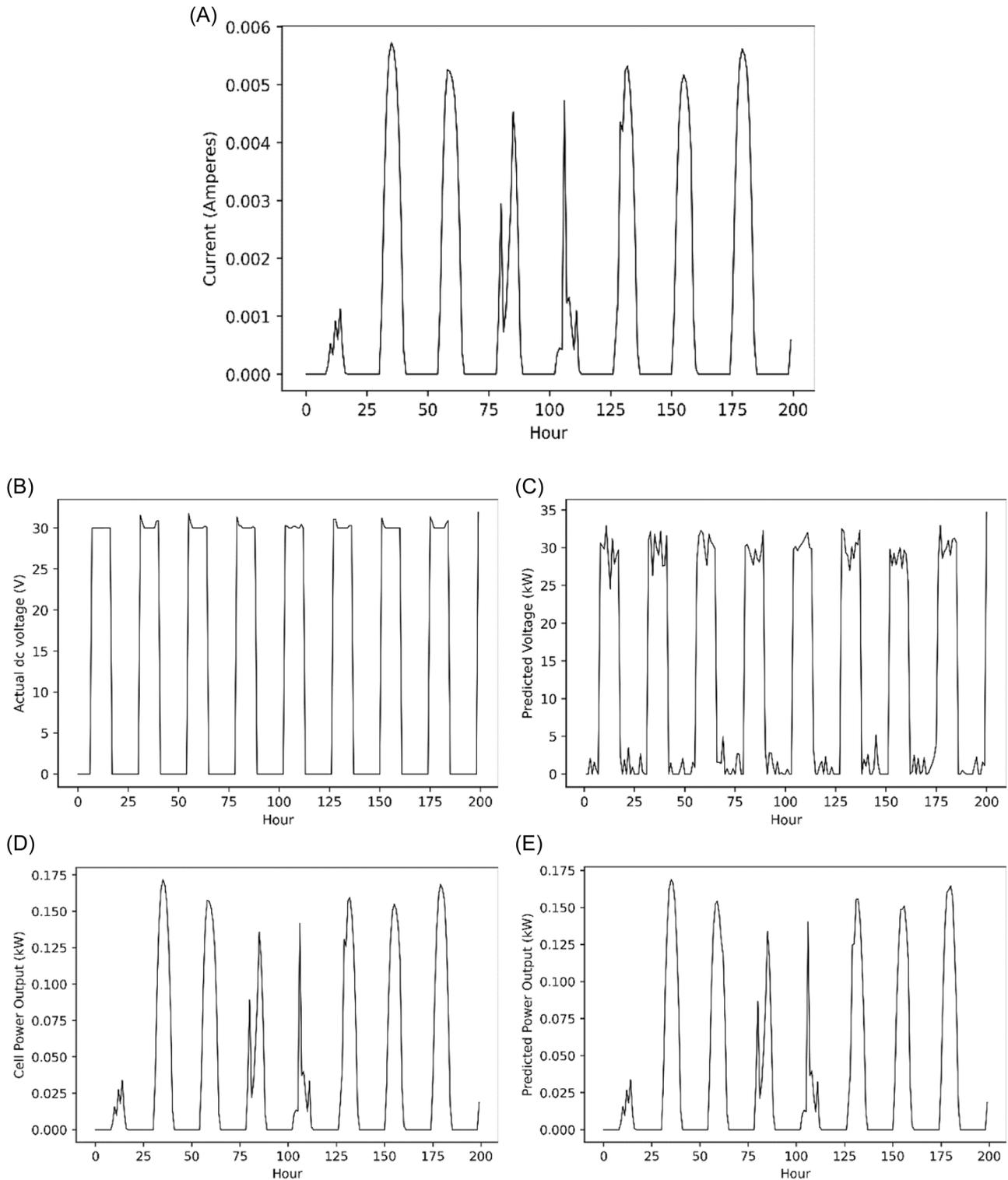

**FIGURE 5** Comparison of ground truth and predicted data (voltage and power).

The model output simulated in real-time using a real-time digital simulator matches the predicted output. An exemplary comparison between predicted output and the output plotted on an MSO logged through the hardware simulation is shown in Figure 7.

## 5 | DISCUSSION

The comparative results over hyperparameters of the proposed model have been presented in Tables 2–5. It was observed that fine-tuning of four specific hyperparameters

**TABLE 2** Comparison of model dimension $d_{eff}$ for the proposed model.

| $d_{eff}$ | $d_{adj}$ | Voltage MAP error (%) | MPPT model average efficiency (%) |
|---|---|---|---|
| 32 | 68 | 5.92 | 95.98 |
| 64 | 100 | 1.21 | 98.32 |
| 128 | 164 | 0.98 | 99.04 |
| 256 | 292 | 0.47 | 99.54 |
| 512 | 548 | 0.49 | 99.21 |

Abbreviations: MAP, mean absolute percentage; MPPT, maximum power point tracking.

**TABLE 3** Comparison of the feed-forward dimension size for the proposed model.

| Feed-forward dimension size | Voltage MAP error (%) | MPPT model average efficiency (%) |
|---|---|---|
| 128 | 1.05 | 98.68 |
| 256 | 0.47 | 99.54 |
| 512 | 1.87 | 97.53 |

Abbreviations: MAP, mean absolute percentage; MPPT, maximum power point tracking.

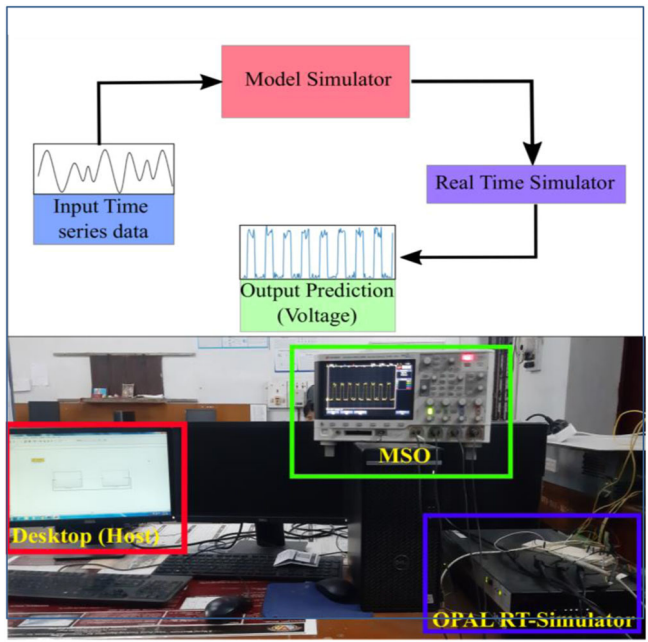

**FIGURE 6** Real-time simulation of the proposed model. MSO, mixed signal oscilloscope

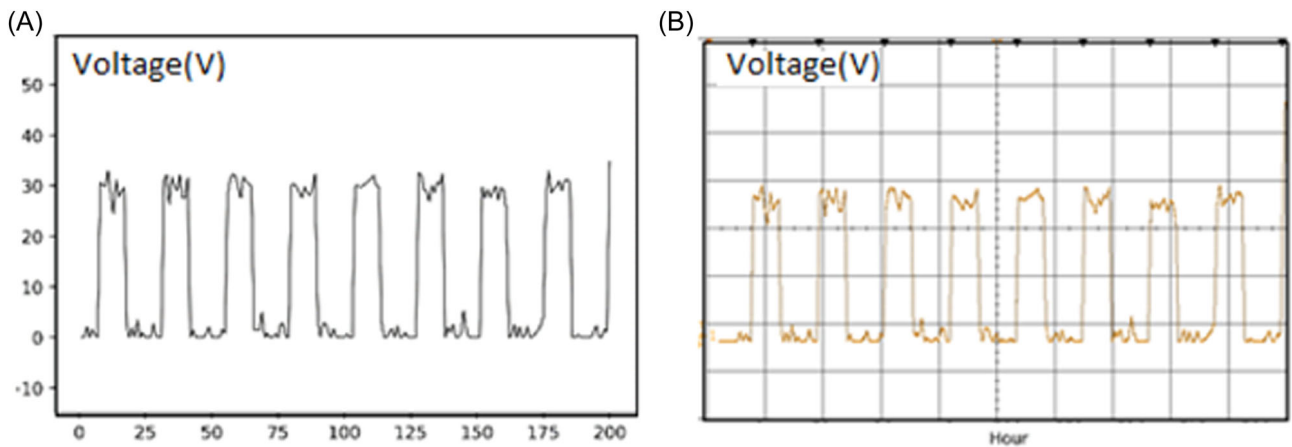

**FIGURE 7** (A) predicted voltage curve, (B) real-time validated voltage curve.

**TABLE 4** Comparison of the dropout probability for the proposed model.

| Dropout probability $p$ | Voltage MAP error (%) | MPPT model average efficiency (%) |
|---|---|---|
| 0.1 | 0.64 | 98.25 |
| 0.2 | 0.47 | 99.54 |
| 0.3 | 3.78 | 96.40 |
| 0.4 | 4.91 | 96.19 |
| 0.5 | 7.04 | 91.74 |

Abbreviations: MAP, mean absolute percentage; MPPT, maximum power point tracking.

**TABLE 5** Comparison of the number of attention heads for the proposed model.

| Number of attention heads | Voltage MAP error (%) | MPPT model average efficiency (%) |
|---|---|---|
| 4 | 1.98 | 97.42 |
| 8 | 1.25 | 98.20 |
| 16 | 0.47 | 99.54 |
| 32 | 0.55 | 99.16 |

Abbreviations: MAP, mean absolute percentage; MPPT, maximum power point tracking.

demonstrated better performance. These hyperparameters are the model dimension $d_{\text{eff}}$, dropout probability $p$, feed-forward dimension size, and the number of heads in the attention layer. For evaluation of the hyperparameter values, the voltage MAP error and the mean MPPT efficiency on the test dataset are chosen. Tweaking the other hyperparameters did not have a significant impact on the performance of the model.

This analysis concludes that the optimal values for the particular MPPT task is 256 for the model dimension size, 256 for the feed-forward dimension size, 0.2 for the dropout probability rate, and 16 for the number of attention heads.

Overall, the low prediction error proves the robustness and effectiveness of modeling the MPPT algorithm as a time series prediction problem. The use of proposed transformer-based encoder proved to be superior over traditional fully connected ANN architectures used in previous works. Also, the use of a comprehensive dataset along with proper feature engineering and representation demonstrates a strong improvement in MPPT tracking. On analysis of the prediction graphs, one observes a very strong correlation between the ground truth and predicted power. There is however a small ripple when the ground truth power is 0. This can be expected from a neural network model since the model is essentially a continuous function, and there will be a change in output with even the slightest change in the input values. However, by judging the nature of the

**TABLE 6** Comparison between papers on MPPT based on neural network-based techniques.

| Reference | Description | Inputs | MPPT model peak efficiency (%) |
|---|---|---|---|
| 56 | Artificial neural network combined with hill climbing | $I_{\text{PV}}$ | 91.04 |
| 17 | Deep Q-network and deterministic policy gradients (DDPG) based reinforcement learning | $I_{\text{PV}}, V_{\text{PV}}, D, \Delta D$ | 93.65 |
| 12 | Fuzzy neural network | $G, T$ | 96 |
| 57 | ANFIS-based MPPT controller | $\Delta P_{\text{PV}}, \Delta I_{\text{PV}}$ | 98.16 |
| 58 | ADALINE neural network trained with real-time recurrent learning (RTRL) | $I_{\text{PV}}, V_{\text{PV}}$ | 98.17 |
| 59 | RBF neural network-based backstepping terminal sliding mode MPPT control | $I_{\text{PV}}, V_{\text{PV}}$ | 98.74 |
| 14 | Bald eagle search (BES) optimized recurrent neural network | $G, T$ | 99.52 |
| 18 | Recurrent neural network-based MPPT | $G, T$ | 99.58 |
| 60 | Artificial neural network with CUK converter topology | $G, T$ | 99.70 |
| 61 | ANFIS trained through a modified firefly algorithm | $T$ | 99.95 |
| Proposed model | Time series transformer | $T$, DNI, DHI, $\alpha_s, \gamma_s, v_w, \phi$, PM, hour, month | 99.98 |

Abbreviations: DHI, diffuse horizontal irradiance; DNI, direct normal irradiance; MPPT, maximum power point tracking.

curve, this issue can be solved by simply putting a high pass filter at a low power point, thus removing the ripple values.

In Table 6, the performance of the proposed model with other neural network-based MPPT approaches presented in the past is compared. Authors have compared the work where a quantitative measure of the performance is reported. Here, peak MPPT efficiency has been considered to compare reported works.

# 6 | CONCLUSION

It is observed that the ambient weather conditions demonstrate cyclic patterns in both the short and long term. As a result, the operating power point of solar PV cells, which is directly affected by ambient weather conditions, also demonstrates cyclic patterns. To effectively model these relations, MPP prediction is modeled as a time series task in this paper. A novel set of data features that comprehensively represent the weather at a given point in time has been proposed. In addition, two time-based features, representing the hour of the day and the month of the year are proposed. These features represent daily cyclic patterns and seasonal cyclic patterns, respectively, and help the model reinforce repeating temporal values. A transformer-based architecture is used to predict the time series values based on the weather data from a given location at a given time. The model achieved a mean prediction error of 0.47% on the test dataset, resulting in an average power efficiency of 99.54% and a peak efficiency of 99.98%, which is significantly better than MPPT performance reported by previous models including neural network-based approaches.

The dataset was sampled from various locations characterized by a diverse range of climatic conditions. The obtained results demonstrate that the proposed technique is capable of modeling various kinds of weather patterns simultaneously and achieving strong and robust predictions. The features proposed and used in this paper prove to be comprehensive methods to represent the local weather conditions. This technique demonstrates proof of concept that it can be applied to various conditions of operation, including different weather conditions, different shading conditions, and variations of other operational parameters. As potential research directions for future works in this domain, multiple data sources may be combined to generate an augmented MPPT dataset containing comprehensive data features representing the true weather conditions. Pure deep learning approaches may also be combined with statistical modeling of time series data (e.g., ARIMA and SARIMA) to give more robust results.

# NOMENCLATURE

| | |
|---|---|
| $I_{PV}$ | PV current |
| $\phi$ | humidity |
| $V_{PV}$ | PV voltage |
| $D$ | duty cycle |
| $\Delta D$ | perturbation |
| $G$ | irradiance |
| $T$ | temperature |
| $\Delta P_{PV}$ | power variation |
| $\Delta I_{PV}$ | current variation |
| $\alpha_s$ | solar altitude angle |
| $\gamma_s$ | solar azimuth angle |
| RL | reinforcement learning |
| MPPT | maximum power point tracking |
| $v_w$ | wind speed |
| PV | photovoltaic |
| MSO | mixed scope oscillator |
| PM | particulate matter (pollution) |
| hour | hour of the day (discretized) |
| month | month of the year (discretized) |
| DNI | direct normal irradiance |
| DHI | diffuse horizontal irradiance |
| P&O | perturb and observe |
| ANN | Artificial Neural Network |

## ORCID

*Palaash Agrawal*  http://orcid.org/0000-0001-5902-585X

*Hari Om Bansal*  http://orcid.org/0000-0001-5835-431X

*Aditya R. Gautam*  http://orcid.org/0000-0001-8732-6158

*Baseem Khan*  http://orcid.org/0000-0002-0562-0933

---

**How to cite this article:** Agrawal P, Bansal HO, Gautam AR, Mahela OP, Khan B. Transformer-based time series prediction of the maximum power point for solar photovoltaic cells. *Energy Sci Eng*. 2022;10:3397-3410. doi:10.1002/ese3.1226